\documentclass[english,aps, superscriptaddress, a4, 12pt, showpacs,nofootinbib]{revtex4}

\usepackage{graphicx}

\usepackage{amssymb}
\usepackage{amsmath}
\usepackage{babel}

\usepackage{color}

\newcommand{\be}{\begin{eqnarray}}
\newcommand{\ee}{\end{eqnarray}}
\newcommand{\nn}{\nonumber }

\usepackage{babel}
\makeatother

\begin{document}

\title{On the scaling behavior of the chiral phase transition in QCD\\ in finite and infinite volume}
\author{Jens~Braun}
\affiliation{Theoretisch-Physikalisches Institut, Friedrich-Schiller-Universit\"at Jena, D-07743 Jena, Germany}
\author{Bertram~Klein}
\affiliation{Physik Department, Technische Universit\"at M\"unchen, James-Franck-Strasse 1, 85747 Garching, Germany}
\author{Piotr~Piasecki}
\affiliation{Theoretisch-Physikalisches Institut, Friedrich-Schiller-Universit\"at Jena, D-07743 Jena, Germany}

\date{\today}                   

\begin{abstract}
We study the scaling behavior of the two-flavor chiral phase transition using an effective quark-meson model. We investigate the transition between 
infinite-volume and finite-volume scaling behavior when the system is placed in a finite box. We can estimate effects that the finite volume and the 
explicit symmetry breaking by the current quark masses have on the scaling behavior which is observed in full QCD lattice simulations. 
The model allows us to explore large quark masses as well as the chiral limit in a wide range of volumes, and extract information about the scaling regimes.
In particular, we find large scaling deviations for physical pion masses and significant finite volume effects for pion masses that are used in current 
lattice simulations.
\end{abstract}


\maketitle

\section{Introduction}

The order and the exact nature of the chiral phase transitions in QCD is important for our understanding of 
the dynamics at the early stages of the universe and of heavy-ion collisions, see 
e.~g.~\cite{BraunMunzinger:2003zd,Stephanov:2008qz}. However, the order and nature of this phase transition
has proven to be notoriously 
difficult to pin 
down~\cite{Karsch:1994hm,Bernard:1999fv,AliKhan:2000iz,D'Elia:2005bv,Philipsen:2005mj,Kogut:2006gt,Cheng:2006qk,Aoki:2006br,Cossu:2007mn,Ejiri:2009ac,Bazavov:2009mi,Borsanyi:2010bp}. 
It has long been surmised, from symmetry arguments and a renormalization group analysis~\cite{Pisarski:1983ms}, 
that the phase transition in QCD at physical quark masses is a crossover and in the domain of a critical 
fixed point of a theory with $O(4)$ symmetry in $d=3$ dimensions. In the limit of $N_f=2$ massless quark 
flavors, a second-order phase transition governed by this fixed point is expected, whereas the phase 
transition becomes first order for $N_f=3$ massless flavors. 

Recently significant progress has been made towards establishing the scaling behavior for the 
two-flavor transition \cite{D'Elia:2005bv,Cossu:2007mn,Kogut:2006gt} and the $2+1$-flavor transition 
with one heavy quark flavor \cite{Ejiri:2009ac}. The transitions for both theories are 
expected to fall into the $O(4)$ universality class.  

Since the chiral flavor symmetry is broken explicitly by the quark masses, it is difficult to directly 
observe the critical behavior in lattice simulations. The transition becomes a crossover and the behavior 
is only pseudo-critical. The actual critical behavior can therefore only be ascertained by means of a scaling analysis.

Because of the similar numerical values of the critical exponents for $O(N)$ models, see e.~g. 
\cite{Ballesteros:1996bd,Grassberger:1997,Guida:1998bx,Hasenbusch:2000ph,Litim:2001hk,Arnold:2001mu,Prokof'ev:2001zz,Litim:2002cf,Kastening:2003iu,Campostrini:2006ms,Benitez:2009xg}, 
it is also difficult to distinguish the expected $O(4)$ scaling behavior 
from $O(2)$ scaling behavior, which might apply due to a residual symmetry in the staggered implementation of 
fermions in lattice QCD. In this context it is useful to also compare results from simulations to scaling functions 
from the appropriate models~\cite{Toussaint:1996qr}.

Scaling in the $O(4)$-model \cite{Toussaint:1996qr,Engels:1999wf,ParisenToldin:2003hq,Engels:2003nq,Braun:2007td,Engels:2009tv} and in quark-meson 
models with $O(4)$ symmetry \cite{Berges:1997eu,Schaefer:1999em, Bohr:2000gp, Stokic:2009uv} in infinite 
volume have been investigated previously using various methods. These studies provide results 
for the scaling functions in infinite volume, which can be used for comparisons. Most of these investigations 
have focussed on the scaling regime close to the critical point, with extremely small amounts of explicit 
symmetry breaking, and have not investigated how this scaling behavior is connected to the behavior at the physical 
pion mass. In order to assess the feasibility of a scaling analysis on the lattice at 
physical values of the pion mass, it appears useful to perform such an investigation in the 
context of a model for dynamical chiral symmetry breaking, e. g. NJL-type models. In such models physical pion masses 
and the chiral limit are both readily accessible.

Universality arguments for the scaling functions do not, in general, allow us to make observations away 
from the critical region where the universal behavior obtains. They are only applicable where the physics of 
the systems are dominated by the critical long-range correlations. Probing shorter distance scales away from the 
critical point, aspects of the underlying short-range physics once again emerge, and different systems once 
again behave differently, even when they are in the same universality class concerning the critical behavior. 
However, the size of the 'critical regime' depends on the additional scales which come into play in different theories. 
For example, in NJL-type models as well as in QCD, there is a gap between the pions as the light Goldstone-modes, and the 
next heavier excitation. This makes e.g. chiral perturbation theory viable as a low-energy effective theory of QCD. 
For this reason, we expect that for values of the pion mass below the physical values and temperatures much smaller
than the critical temperature, differences between the behavior of the models and QCD ought to remain small, so that the 
model results still have relevance for our expectations about full QCD. In the vicinity of the critical temperature
this assumption may no longer hold since at the transition lots of higher resonances come into play~\cite{Karsch:2003zq}. 
In fact, the dependence of the critical temperature on the pion mass is much stronger in NJL-type models than that which 
has been found in lattice simulations. We shall address this issue below.

In addition to the issue of finite pion masses, lattice simulations are also always performed in finite simulation volumes. 
Thus one has to take care to exclude finite-volume 
effects in order to observe the scaling behavior expected for infinite-volume systems. 
Such finite-volume effects can also affect the phase transition at finite density \cite{Palhares:2009tf,Abreu:2009zt,Braun:2010pp}.
Alternatively, one can turn the appearance 
of finite-size effects into an advantage and perform a finite-size scaling analysis~\cite{Fisher:1971ks,Fisher:1972zza} to obtain additional 
information about the scaling behavior \cite{D'Elia:2005bv,Cossu:2007mn,Engels:2005te,Engels:2001bq,Braun:2008sg,Kogut:2006gt}. 
Finite-Size scaling functions relevant for the analysis of the chiral order parameter have been obtained e.g. in \cite{Engels:2001bq,Braun:2008sg}. 

We can distinguish between two regimes, one where the scaling behavior conforms to that for infinite volume, and one where one
observes clear finite-size scaling effects. In between these two regimes, it is not obvious how the finite volume 
will affect the scaling behavior, and an investigation of the transition between these regimes in a model system is worthwhile.

A crucial part of the underlying argument for our expectations of the scaling behavior at the chiral phase transition 
in QCD relies on the idea of dimensional reduction: If we consider a system in infinite volume, but at finite temperature, 
as a system in Euclidean space, close to a critical point, then the wavelength of the critical fluctuations eventually 
become larger than the extent of the system in the Euclidean time direction. This leads to an effectively three-dimensional 
system, and subsequently to our expectation of three-dimensional scaling behavior in the $O(4)$ universality class 
for two-flavor QCD. 

In order to observe infinite-volume scaling behavior in the expected universality class in finite-volume lattice 
simulations, on the one hand the extent in Euclidean time direction has to be small enough to lead to dimensional reduction. 
On the other hand the extent in the spatial directions has to be large enough to minimize finite-volume effects. The scale 
set by the temperature must then be compared to that set by the long-range correlations, which is bounded by the pion mass. 
This is a potential issue for lattice simulations, where the overall scale for the lattice spacing is set by the value 
of the gauge coupling, but the aspect ratio of the lattice in spatial and Euclidean time direction remains fixed. It appears 
therefore worthwhile to investigate both the infinite-volume scaling behavior and the deviations from this behavior for fixed 
aspect ratio in a finite volume, both for very small pion masses close to the chiral limit, and for physical values.

Considering that the physical pion mass is around $140$ MeV, and the corresponding chiral phase transition temperature is 
approximately $150-180$ MeV~\cite{Cheng:2006qk,Aoki:2006br,Bazavov:2009mi,Borsanyi:2010bp}, 
the proposition that the system should experience dimensional reduction for a physical 
choice of parameters does not appear \emph{a priori} obvious. 

The scaling analysis presented in this work may help to shed light on the scaling analysis in actual QCD studies since the physics 
of the chiral phase transition is mainly determined by the long-range effective degrees of freedom at low momentum scales, namely 
the pions as the Goldstone modes of chiral symmetry breaking. As far as these degrees of freedom determine the behavior of the transition, 
according to universality arguments, the much simpler model system should exhibit the same critical behavior as QCD.
Of course, our model approach cannot answer questions outside the applicability of the model. For example, the order of the phase transition 
is in our case already fixed by the $O(4)$-symmetry of the model, while the order of the transition in QCD has not yet been unambiguously
determined \cite{Bernard:1999fv,D'Elia:2005bv, Philipsen:2005mj}.  Moreover, we must also limit our investigation to the chiral phase transition.
On the other hand, in calculations based on renormalization group methods the model has been used to investigate chiral properties of QCD beyond 
the mean-field approximation, 
such as the critical behavior and the quark mass dependence of the chiral transition~\cite{Berges:1997eu,Schaefer:1999em,Braun:2005fj} 
as well as the critical behavior at finite density~\cite{Schaefer:2004en}. In the past few years it has also been combined with Polyakov loop results 
from lattice QCD simulations to improve the description of thermodynamical observables, see e.~g. 
Refs.~\cite{Meisinger:1995ih,Pisarski:2000eq,Fukushima:2003fw,Ratti:2005jh,Megias:2004hj,Ghosh:2006qh,Sasaki:2006ww}. Corrections
beyond the mean-field approximation in the (P)NJL/(P)QM model have been considered in \cite{Schaefer:2007pw,Cristoforetti:2010sn,Skokov:2010wb,Herbst:2010rf}. 

The paper is organized as follows: In Sect.~\ref{sec:FV_RG} we briefly discuss the model and our RG approach to a scaling analyis
in infinite and finite volumes. In Sect.~\ref{sec:IV} we present the results from our scaling analysis in infinite volume while we discuss
scaling in finite volumes in Sect.~\ref{sec:FV}. Our conclusions are summarized in Sect.~\ref{sec:conc}.

\section{Chiral Model and non-perturbative RG approach}\label{sec:FV_RG}
In this section we discuss our RG  approach to a finite-volume scaling study of the chiral phase transition in QCD. We briefly introduce 
the (linear) quark-meson model and discuss the derivation of the flow equations for finite and infinite volume studies. 
A detailed discussion of the derivation and the approximations  involved can be found in 
Refs.~\cite{Braun:2007td,Braun:2004yk,Braun:2005gy,Braun:2005fj}.

To study the chiral phase transition and its scaling behavior for infinite and finite volume, we employ the chiral quark-meson 
model. This model is an $O(4)$-invariant linear $\sigma$-model with $N_f=2$ quark flavors with $N_c=3$ colors and $N_{f}^{2}=4$ 
mesonic degrees of freedom. The mesons are coupled to the (constituent) quarks in an $SU(2)_{L}\times SU(2)_{R}$ 
invariant way. We stress that it is an effective low-energy model for dynamical spontaneous chiral symmetry breaking at 
intermediate scales. However, it does not contain gluonic degrees of 
freedom and is not confining. 

At the scale $\Lambda$, the quark-meson model is defined by the bare effective action 
\begin{eqnarray} 
  \Gamma_{\Lambda}[\bar q,q,\phi]&=& \int d^{4}x \Big\{
  \bar{\Psi} \left({\partial}\!\!\!\slash + 
  g(\sigma+i\vec{\tau}\cdot\vec{\pi}\gamma_{5})\right)\Psi
  +\frac{1}{2}(\partial_{\mu}\phi)^{2}+U_{\Lambda}(\phi^2) - H \sigma \Big\} 
\label{eq:QM}
\end{eqnarray} 
with $\phi^{\mathrm{{T}}}=(\sigma,\vec{\pi})$ and $\Psi$, $\bar{\Psi}$ denote the fermion spinors associated with the quark fields.
We choose the first component of the vector $\phi$ to be the radial mode associated with the $\sigma$ meson. Note that due to the explicit 
symmetry breaking the ground state of the theory is only symmetric under $O(3)$ transformations.
The mesonic potential at the ultraviolet (UV) scale is characterized by two couplings, $m^2_\Lambda$ and $\lambda_\Lambda$,
\begin{equation}
  \label{eq:pot_UV} 
  U_\Lambda(\phi^{2}) =
  \frac{1}{2}m_\Lambda^{2}\phi^{2} +
  \frac{1}{4}\lambda_\Lambda(\phi^{2})^{2}
  \,.
\end{equation}
The quarks and the mesons are coupled via chirally symmetric Yukawa term with $g$ being the coupling.
The linear term in $\sigma$ results from a bosonization of the current quark mass term $\sim \bar{\Psi} m_c\Psi$ and
the symmetry breaking parameter~$H$ is therefore related to the current quark mass 
and $m(\Lambda)=m_{\Lambda}$: $H=m_c m_{\Lambda}^2/g$.
We study the quark-meson model in the so-called local potential approximation (LPA), where we neglect a possible space dependence 
of the expectation value $\langle\phi\rangle$ and take the wave-function renormalizations $Z_{\phi}$ and $Z_{\psi}$
to be constant, $Z_{\phi}=1$ and \mbox{$Z_{\psi}=1$}. This approximation should not be confused with a mean-field approximation. In fact,
our approximation includes already beyond mean-field effects. A detailed discussion of the relation of the present approximation (LPA)
to the mean-field approximation can be found in Refs.~\cite{Braun:2008pi,Braun:2009si}. At finite temperature we moreover 
neglect a possible difference of the wave-function renormalization parallel and perpendicular to the heat-bath. 
It has been found that the latter approximation does not strongly affect the 
dynamics near the phase transition~\cite{Braun:2009si}. Since the anomalous dimensions associated with $Z_{\phi}$ and $Z_{\psi}$ are small compared 
to one, see e. g. Refs.~\cite{Tetradis:1993ts,Berges:1997eu,Braun:2009si}, our approximation, in which the running of the wave-function renormalization is 
neglected, is justified for a study of finite-size scaling. In fact, corrections beyond the local potential approximation changes the resulting
critical exponents only at the one-percent level~\cite{Tetradis:1993ts,Berges:1997eu,Benitez:2009xg}. For products of critical exponents
that enter our scaling analysis the changes are slightly smaller.

For our derivation of the RG flow equation for the effective action we employ the Wetterich equation~\cite{Wetterich:1992yh}:
\be
\partial _t \Gamma_k =\frac{1}{2}\text{STr}\,(\partial_t R_k)\cdot
\left[\Gamma_k ^{(2)}+R_k\right]^{-1}\,,\label{eq:FlowEq}
\ee
where the dimensionless flow variable $t$ is given by $t=\ln (k/\Lambda)$. Reviews of and introductions to functional RG approaches
can be found, e. g., 
in Refs.~\cite{Litim:1998nf,Bagnuls:2000ae,Berges:2000ew,Polonyi:2001se,Delamotte:2003dw,Pawlowski:2005xe,Gies:2006wv,Delamotte:2007pf,Sonoda:2007av,Rosten:2010vm}.
The regulator function $R_k$ specifies the details of the Wilsonian momentum-shell
integrations and has to satisfy certain constraints~\cite{Wetterich:1992yh}. Since the choice of the 
regulator function is at our disposal, we can use it to optimize the RG flow~\cite{Litim:2000ci,Litim:2001fd,Litim:2001up,Pawlowski:2005xe}.
In the following, we employ~\cite{Litim:2001up} 
\be
R_k(\vec{p}^{\,2})=\vec{p}^{\,2} r(\vec{p}^{\,2}/k^2)\qquad\text{with}\qquad r(x)=\left(\frac{1}{x}-1\right)\Theta(1-x)\,.\label{eq:regulator}
\ee
In order to derive the RG flow equations for a system in a finite four-dimensional Euclidean volume $L^{3}\times 1/T$ with temperature
$T$, we replace the continuous momenta by discrete momenta and correspondingly each momentum integral in the evaluation of the 
trace in Eq.~\eqref{eq:FlowEq} by a sum:
\begin{equation} 
 \vec{p}^{\,2}\rightarrow 4\pi^2 \vec{n}^2 \equiv 4\pi^2 (n_1^2 + n_2^2 +n_3^2)\,\quad\text{and}\quad
  \int_{-\infty}^\infty d^3 p \rightarrow
  \left(\frac{2\pi}{L}\right)^3 \sum_{n_1 = -\infty}^{\infty}\sum_{n_2 = -\infty}^{\infty}\sum_{n_3 = -\infty}^{\infty}\,. 
\end{equation} 
Since we are ultimately interested in a comparison to scaling behavior of the chiral order parameter as measured in lattice QCD
simulations, see e.~g.~Ref.~\cite{Ejiri:2009ac}, we choose periodic boundary conditions for the bosons and fermions in the spatial directions.

For studying scaling behavior it is convenient to deal with dimensionless quantities rather than dimensionful
quantities. Therefore we introduce the dimensionless potential $u$,
the dimensionless fields $\varphi$ and the dimensionless symmetry breaking parameter $c$ by
\be
u_k =k^{-4} U_k \,,\quad \varphi_i=k^{-1}\phi_i\,,
\quad\text{and}\quad c=k^{-3}H\,.
\label{eq:dimquant}
\ee
Note that the Yukawa coupling in $d=4$ is already a dimensionless quantity. With these definitions the flow equation for the effective potential 
of the quark-meson model in a finite box with length $L$ at finite temperature $T$ is then given by
\begin{widetext}
  \begin{eqnarray}
    \label{eq:fv_ft_flow_equation}
    \partial_t u_k(\varphi^2)
    &=& -4 u_k + {\mathcal B}(kL)\Bigg[ \frac{3}{\epsilon_\pi} 
    \left( \frac{1}{2}+n_B(\epsilon_\pi) \right)
    +\frac{1}{E_\sigma} \left(
      \frac{1}{2}+n_B(\epsilon_\sigma)\right) \nn\\
   &&\quad\quad\quad\quad\quad\quad\quad\quad -\frac{2 N_c N_f}{\epsilon_q} \Big( 1-n_F(\epsilon_q,\mu)- n_F(\epsilon_q,-\mu)
    \Big) \Bigg] \, ,
  \end{eqnarray}
\end{widetext}
where the first two terms correspond to contributions of the mesonic modes, and the last term with opposite overall sign 
corresponds to the quark contributions.   The effective energies are given by
\begin{equation}
\epsilon_i=\sqrt{1+m_i^2} \; , \quad i \in \{ \pi, \sigma, q \} \, , 
\end{equation}
with 
\begin{eqnarray} 
  m_{\pi}^2 = 2 \frac{\partial
    u_k}{\partial \varphi^2}\,,\qquad
  m_{\sigma}^2 = 2 \frac{\partial u_k}{\partial \varphi^2} + 4 \varphi^2
  \frac{\partial^2 u_k}{\partial (\varphi^2)^2}\,,\qquad m_q^2 = g^2 \varphi^2\,.
\end{eqnarray} 
The temperature dependence of the RG flow of the effective potential is governed by the 
bosonic and fermionic distribution functions $n_B$ and $n_F$, respectively:
\begin{equation}
  n_B(\epsilon)=\frac{1}{e^{\epsilon/\tilde{t}}-1} \; , \quad n_F(\epsilon,\mu)=\frac{1}{e^{
      (\epsilon-\mu)/{\tilde{t}}}+1} \,,
\end{equation}
where $\tilde{t}=T/k$ denotes the dimensionless temperature. The dependence on the finite spatial volumes is encoded in the
function ${\mathcal B}$: 
\begin{equation} 
  {\mathcal B}(kL)=\frac{1}{(kL)^3} \sum_{n_1 = -\infty}^{\infty}\sum_{n_2 = -\infty}^{\infty}\sum_{n_3 = -\infty}^{\infty} \Theta\!\left( (kL)^2    
    -4\pi^2 \vec{n}^2\right). 
\end{equation} 
The asymptotic behavior of this function for small and large dimensionless box sizes $kL$ is given by
\begin{eqnarray} 
\lim_{kL\to 0} {\mathcal B}(kL) \sim \frac{1}{(kL)^3}\,\qquad\text{and}\qquad
\lim_{kL\to \infty} {\mathcal B}(kL) = \frac{1}{6\pi^2}\ . 
\end{eqnarray} 
The behavior for small $kL$ reflects the fact that the dynamics of the system is mainly governed by the spatial zero modes in this limit.
On the other hand we recover the same flow equation as found in Refs.~\cite{Braun:2003ii,Schaefer:2004en,Stokic:2009uv}
for $kL\to\infty$.

In order to solve the RG flow for the scale-dependent effective mesonic potential $u_k$, we expand the potential in a Taylor series in
scale dependent local $n$-point couplings $a_{n, k}$ around its scale
dependent minimum $\langle \varphi_0 \rangle$ 
\begin{eqnarray}
  \label{eq:pot_ansatz}
  u_k(\varphi^2) &=&  \sum_{n=0}^{N_{\text{max}}}\frac{a_{n,k}}{2^n n!}
  ( \varphi^2\!-\!  \langle \varphi_0\rangle^2)^{n}.
\end{eqnarray} 
The presence of the symmetry-breaking term $-H\sigma$ in our ansatz~\eqref{eq:QM} induces a shift of the minimum from its value in the chiral 
limit. Following Ref.~\cite{Braun:2008sg}, the condition
\begin{equation} 
  \label{eq:min_cond}
  \frac{\partial}{\partial
    \varphi_0} u_k(\varphi^2) \Bigg|_{\varphi_0=\langle \varphi_0\rangle,\varphi_i= 0} \stackrel{!}{=} c
\end{equation} 
ensures that the potential is always expanded around the actual physical minimum.
From Eq.~\eqref{eq:min_cond} it follows that the RG flow of the coupling $a_{1, k}$ and the minimum $\langle\varphi_0\rangle$ 
are related by the simple condition
\begin{equation}
  \label{eq:min_cond2}
  a_{1,k} \langle \varphi_0\rangle = c\,. 
\end{equation} 
This condition keeps the potential minimum at $\varphi =(\langle\varphi_0\rangle, \vec{0})$.

The RG flow equations for the couplings $a_{n,k}$ and $\langle\varphi_{0}\rangle$ can be obtained by expanding the equation for the effective potential,
Eq.~\eqref{eq:fv_ft_flow_equation}, around the scale-dependent $\langle\varphi_{0}\rangle$ and then projecting it onto the derivative with respect to $k$
of the ansatz, Eq.~\eqref{eq:pot_ansatz}. This procedure results in an infinite set of flow equations for $\langle\varphi_{0}\rangle$ and the couplings 
$a_{n}(k)$. In order to obtain a finite set of flow equations, we truncate the Taylor series,
Eq.~\eqref{eq:pot_ansatz}, at a fixed order $N_{\text{max}}=4$ and include thus fluctuations around the physical ground-state configuration 
up to order $2N_{\text{max}}$ in the fields. Note that such an expansion represents a systematic expansion in $m$-point functions 
$\Gamma ^{(m)}$ where $m=2n$ determines the number of external legs. The quality of such an expansion of the order-parameter potential
in powers of $\varphi^2$ has been studied quantitatively in~Ref.~\cite{Tetradis:1993ts} at vanishing temperature and for the proper-time RG 
in LPA at finite temperature in~Ref.~\cite{Papp:1999he}. Moreover the order-parameter potential has been computed 
in LPA without making use of a Taylor expansion in~$\varphi$ in Ref.~\cite{Schaefer:2004en}

\section{Scaling analysis in the infinite volume limit}\label{sec:IV}
\label{sec:infvolscaling}
In the vicinity of a critical point, where the dynamics of the system are dominated by critical long-range fluctuations, the singular part of the 
free energy density of the system satisfies to leading order the scaling relation
\be
f_s(t, h) =\ell^{-d}  f_s(t \ell^{y_t},h \ell^{y_h} ),  
\ee
where $\ell$ is a dimensionless rescaling factor which can be chosen arbitrarily, and $t=(T-T_{\rm c})/T_0$ and $h=H/H_0$ are the reduced temperature, 
measured from its critical value, and the external symmetry-breaking field, normalized in a suitable way. 

The exponents $y_t$ and $y_t$ specify all critical exponents for the scaling behavior, 
\be
y_t = \frac{1}{\nu}, \quad y_h =\frac{\beta \delta}{\nu}, 
\ee
when taken in combination with the additional scaling relations $\gamma = \beta (\delta -1)$, $\gamma = (2-\eta)\nu$.

As a consequence of the scaling relation, observables such as the order parameter, identified with the pion decay constant $M \equiv f_\pi$ 
in the model, and the susceptibilities $\chi_\pi$ for transverse Goldstone modes and $\chi_\sigma$ for longitudinal modes can be expressed 
in terms of universal scaling functions. By choosing the scaling factor $\ell$ such that either $t\ell^{y_t} = 1$ or $h \ell^{y_h} =1$, 
the free energy density becomes a function of only a single scaling variable, with an explicit dependence on either $t$ or $h$. 
Thermodynamic observables which can be expressed in terms of derivatives of $f(t, h)$ with respect to its arguments can then also be 
expressed in terms of such scaling functions \cite{Widom:1965xx}.

For the order parameter, one finds the scaling relation
\be
M &=& h^{1/\delta} f_M(z), \quad z= t/h^{1/(\beta \delta)}\label{eq:Mz}
\ee
where $z$ is the scaling variable, and $f_M(z)$ is the scaling function normalized to $f_M(0)=1$. Asymptotically for small values 
of $h$ and $t<0$, i.e. for large values for $-z$, we have $f_M(z) \simeq (-z)^\beta$. These two conditions determine the normalization 
constants $T_0$ and $H_0$, such that $M = h^{1/\delta}$ for $t=0$ and $M=(-t)^\beta$ for $h=0$ and $t<0$. 

In the LPA used in this paper, the static susceptibilities are related to the masses of the mesonic 
modes according to
\be
\chi_\pi = \frac{1}{M_\pi^2} \quad \mathrm{and} \quad \chi_\sigma =\frac{1}{M_\sigma^2}   
\ee
for the transverse and the radial modes, respectively. Since the susceptibility for the transverse mode is related to the order 
parameter $M = \langle \varphi_0 \rangle = f_\pi$ according to Eq.~\eqref{eq:min_cond2}, i.e. 
\be
\chi_\pi = \frac{M}{H}=\frac{\langle \varphi_0\rangle}{H},
\ee
the transverse susceptibility does not contain any additional information beyond that contained in the behavior of the order 
parameter, and we will therefore not consider it separately.

The longitudinal susceptibility $\chi_\sigma = \frac{\partial M}{\partial H}$ can be expressed in terms of the scaling 
function $f_\chi(z)$, which is related to the scaling function $f_M(z)$ and its derivative according to
\be
\chi_\sigma &=& \frac{h^{1-1/\delta}}{H_0} f_\chi(z) = \frac{h^{1-1/\delta}}{H_0} \frac{1}{\delta}  \left[ f_M(z) -\frac{z}{\beta} f_M^\prime(z)\right].
\ee
It corresponds to the chiral susceptibility, i.e. to the susceptibility of the chiral condensate $\langle \bar \psi \psi \rangle$ 
with respect to a change in the current quark mass $m_c$. 

We wish to stress that the scaling analysis for a theory with $d=3$ is here performed for a theory in $d=4$ Euclidean 
dimensions, where the temperature (in a field-theoretical sense) is given by the Euclidean time extent of the volume. 
This means that we will in fact only observe scaling in a three-dimensional universality class when the conditions for 
a dimensional reduction are met.
In contrast, many earlier determinations used a three-dimensional theory, where the transition was determined by the critical 
value of one of the couplings of the theory see e.g. \cite{Bohr:2000gp,Braun:2007td} for RG and \cite{Engels:1999wf,Engels:2000xw,Engels:2001bq} 
for spin model results. A scaling analysis in infinite volume based on functional RG approaches, in which temperature has been introduced in 
a field-theoretical sense, has been performed in Refs.~\cite{Berges:1997eu,Bohr:2000gp,Stokic:2009uv}. In this work, we study scaling for small
and large pion masses in infinite but also in finite volumes.
 
In the LPA, the Yukawa coupling $g$, the symmetry breaking field $H$ as well as the expansion coefficients $a_{n,k}$ of the order 
parameter potential $u_k$ are parameters at the UV cutoff scale which are at our disposal. In principle it is possible to fix these parameters
from first principles by employing an RG group approach to full QCD~\cite{Gies:2002hq,Braun:2005uj,Braun:2006jd,Braun:2008pi, Braun:2009gm,Braun:2009ns}. However, 
we shall not follow this strategy in the present paper. Since we are interested in a study of the chiral phase transition in QCD, we fix the 
parameters such that we reproduce the physical values of the pion decay constant, the pion mass and the constituent quark mass
in the infrared (IR) limit, i.~e. for $k\to 0$:
\be
M_{\pi}\approx 138 \,\text{MeV},\qquad f_{\pi}\approx 93\, \text{MeV},\qquad M_q\approx 298\,\text{MeV}.
\ee
Since we use only three IR observables to fix at least three UV parameters, there is some ambiguity in the parameter-fixing procedure. 
However, we have checked that our results depend only weakly on the actual values of the UV parameters of our model, provided a 
given parameter set yields the same IR values for our physical observables. This observation is in accordance with earlier studies, see 
e.~g. Refs.~\cite{Braun:2007td,Braun:2008sg}. To be specific, we choose
\be
g=3.2\,,\qquad a_{1,\Lambda}\equiv\frac{m^2_{\Lambda}}{\Lambda^2}\approx 0.547\,,\qquad a_{2,\Lambda}\equiv 2\lambda_{\Lambda}
\approx 67.2\,,\qquad a_{n,\Lambda}=0\quad \text{for }n\geq 3
\ee
at the scale $\Lambda=3\,\text{GeV}$. We have also checked that our results for, e. g., the critical temperature are 
independent of $\Lambda$ for $\Lambda\gtrsim 3\,\text{GeV}$. For $\Lambda < 3\,\text{GeV}$ we find that our results depend on $\Lambda$. 
Our choice for $\Lambda$ is larger than typical values for the UV cutoff in (P)NJL/(P)QM-type model studies where $\Lambda$ is identified 
with some hadronic scale below which a mesonic description of QCD may become applicable.
A (weak) dependence of, e. g., the critical temperature on $\Lambda$ does not usually play a role in such model studies. In the present study, however,
we have to take care that our results for, e.~g. the critical temperature, do not exhibit a dependence on the UV cutoff since it would spoil our 
scaling analysis. From a phenomenological point of view, choosing a large UV cutoff $\Lambda$ for a low-energy model is not an issue, 
provided the UV parameters of the model have been chosen such that the IR physics remains unchanged. The predictions from the model 
for, e. g., the (chiral) phase boundary, are then not (or only slightly) affected by the choice for the cutoff $\Lambda$.

From now on we leave our choice for the couplings at the initial UV scale unchanged for all temperature and volume sizes. We are then
left with one parameter, namely the external symmetry breaking field $H$, which mimics the current quark mass. Different values of $H$ 
translate directly into different values of the pion mass $M_{\pi}$. For $H\to 0$ (chiral limit) we find the following value for the
critical temperature:
\be
T_{\rm c}\approx 144.949346731961\,\text{MeV}\,.
\ee
Note that in our model study such a high accuracy for $T_{\rm c}$ turns out to be necessary to resolve the 
scaling region in the infinite-volume limit, see below. This has also been observed in scaling studies of $O(N)$ models in
$d=3$, see e. g.~\cite{Braun:2007td,Braun:2008sg}. Studying the model for various values of $H$ in the vicinity of $H=0$, 
we obtain from Eq.~\eqref{eq:Mz} the values for the normalization constants $T_0$ and $H_0$:
\be
T_0 \approx 23.862066412513776\,\text{GeV},\qquad H_0 \approx 346.37722832663883\,\text{GeV}^3\,.
\ee
We find that the critical exponents of our model are consistent with the well-known LPA values found in studies of 
$O(N)$ models~\cite{Litim:2001hk,Braun:2007td,Braun:2008sg}:
\be
\beta\approx 0.4022\,,\qquad \delta \approx 5.0\,.
\ee
In the following we shall use these values for the critical exponents in our scaling analysis.

\begin{figure}[t]
\begin{center}
\includegraphics[scale=0.76]{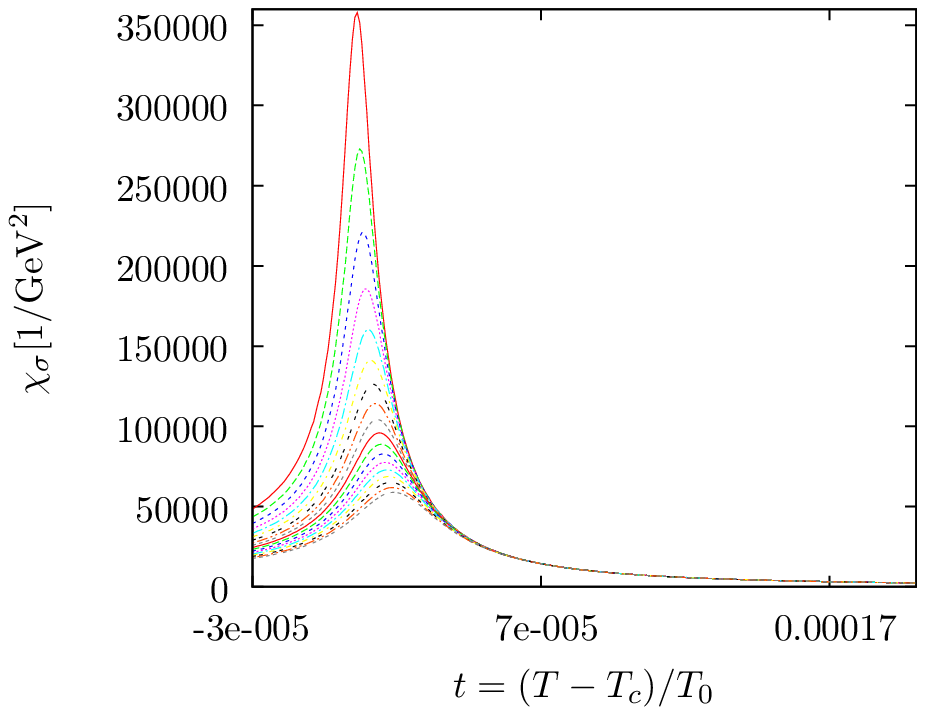}
\includegraphics[scale=0.76]{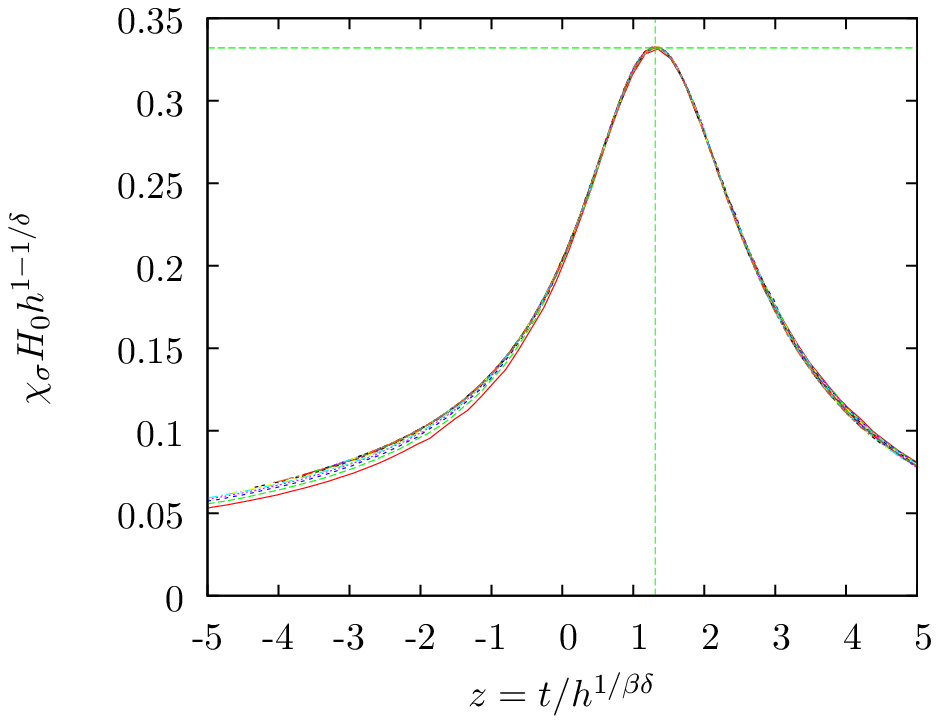}
\caption{Left panel: Chiral susceptibility $\chi_{\sigma}$ as a function of the reduced temperature $t$ for various pion 
masses $M_{\pi}=0.2,\dots,0.9\,\text{MeV}$ (from top to bottom). Right panel: Rescaled chiral susceptibility $\chi_{\sigma}$ as a function 
of the scaling variable $z$ for various pion masses $M_{\pi}=0.2,\dots,0.9\,\text{MeV}$.
\label{fig:susIV}}
\end{center}
\end{figure}

Let us now briefly discuss our results in the infinite-volume limit. In Fig.~\ref{fig:susIV} we show
the chiral susceptibility $\chi_{\sigma}$ and the rescaled chiral susceptibility for various pion masses 
from $M_{\pi}=0.2,\dots, 0.9\,\text{MeV}$. From the results for the chiral susceptibility $\chi_{\sigma}$, 
(left panel) we deduce that the scaling region in our model is indeed very small.
This requires a high accuracy in the determination of $T_{\rm c}$, the critical temperature in the chiral limit. 
Since the susceptibility $\chi_{\sigma}$ is proportional to the squared correlation length, we define the peak of the 
susceptibility $\chi_{\sigma}$ to be the pseudo-critical temperature $T_{\rm p}(H)\equiv T_{\rm p}(M_{\pi})$ which is associated with
long-range correlations. For high temperatures $T\gg T_{\rm c}$ the system is outside of the scaling region and the results for the susceptibility 
$\chi_{\sigma}$ fall onto a single line, $\chi_{\sigma}\sim 1/T^2$. 

After rescaling, the curves for $\chi_{\sigma}$ for $M_{\pi}=0.2,\dots, 0.9\,\text{MeV}$ fall onto a single line 
and are almost indistinguishable at the scale of the plot. The maxima of the curves are located at $z_{\rm p}\approx 1.3155$, 
see also \cite{Braun:2007td,Braun:2008sg}. Thus scaling corrections are bound to be small in this pion mass range. 
However, corrections to scaling become soon apparent for larger values of the pion mass, as we shall see below.

\begin{figure}[t]
\begin{center}
\includegraphics[scale=0.68]{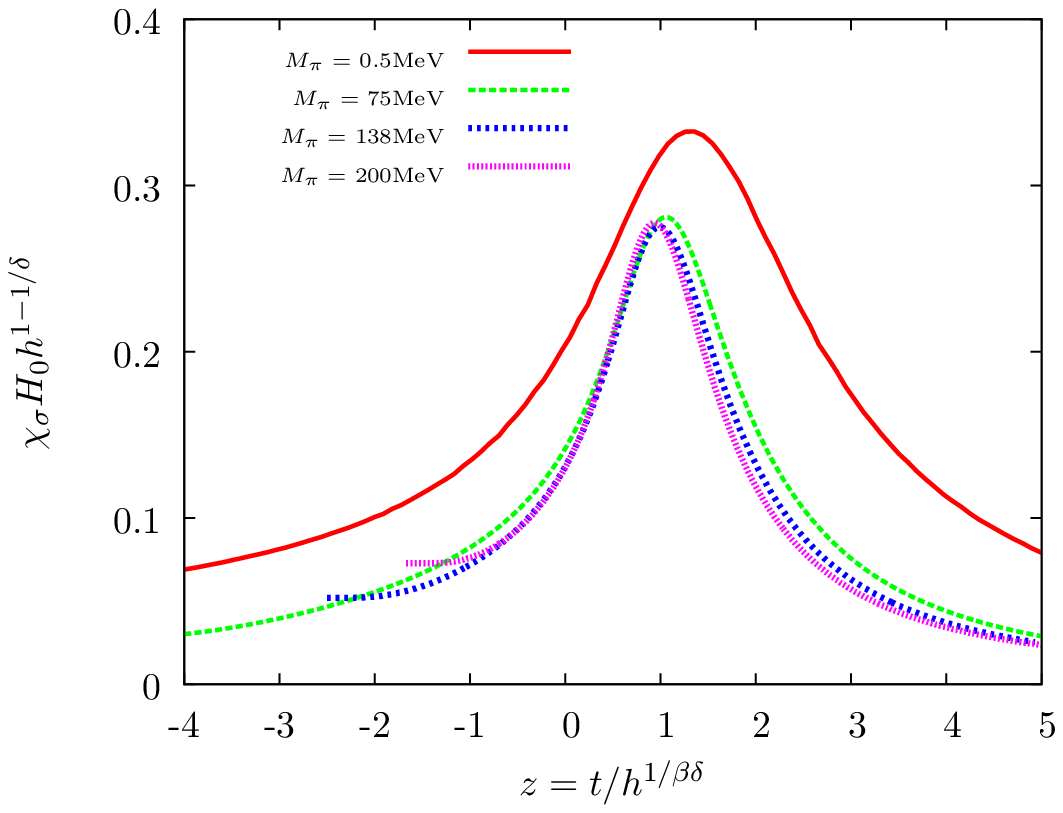}
\includegraphics[scale=0.68]{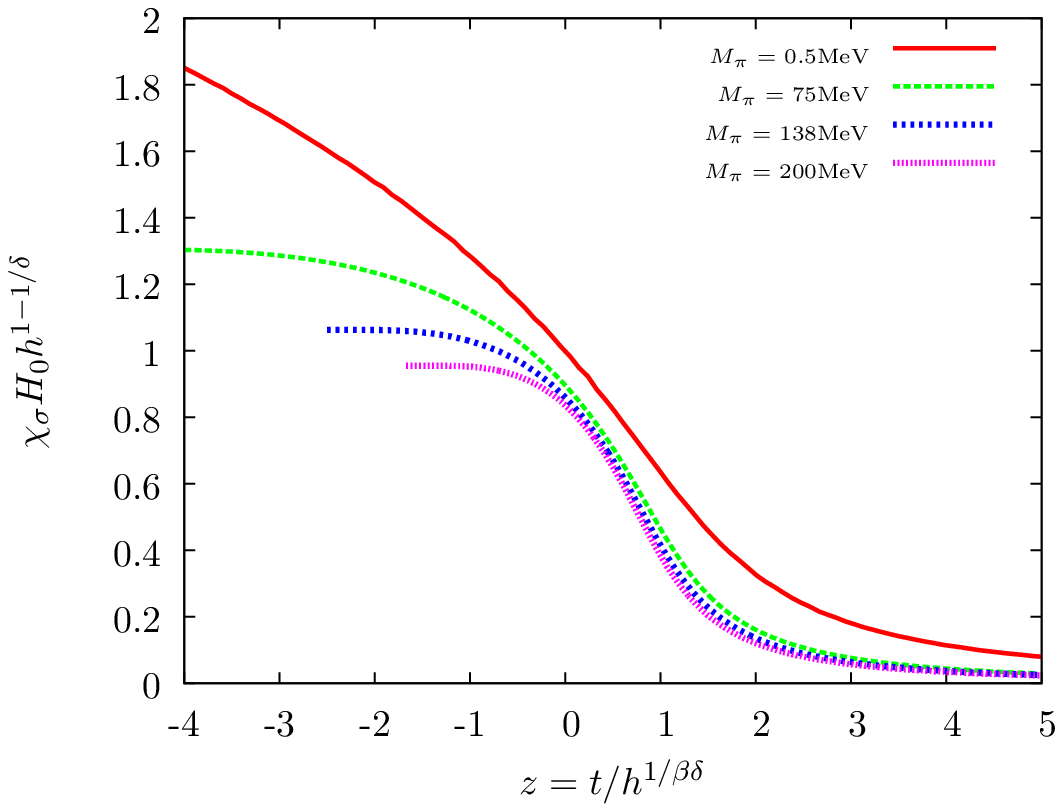}
\caption{Left panel: Rescaled $\sigma$-suszeptibility as a function of $z$ for various pion masses and $L\to\infty$.
Right panel: Rescaled pion decay constant $f_{\pi}$ (order parameter) as a function of $z$ for various pion masses and $L\to\infty$.
\label{fig:susIVmpi}
}
\end{center}
\end{figure}

In Fig.~\ref{fig:susIVmpi} we present our results for the rescaled chiral susceptibility $\chi_{\sigma}$ and the rescaled
pion decay constant as a function of the scaling variable $z$ for $M_{\pi}=0.5\,, 75\,, 138\,, 200\,\text{MeV}$.  
The pion masses used to obtain these results include the physical value as well as the currently smallest value used in lattice 
simulations~\cite{Ejiri:2009ac,Karsch:2010ya}. We observe that the curves for the rescaled susceptibility
and the rescaled order parameter do not fall onto a single line. This ought to be the case if corrections to scaling were small in this
pion mass regime. On the contrary, the rescaled susceptibility and the rescaled order parameter differ significantly from the 
scaling function obtained for small pion masses. On the other hand the results for the rescaled susceptibility and the rescaled order parameter
for $M_{\pi}=75\,, 138\,, 200\,\text{MeV}$ fall almost on one curve for $z\gtrsim -1$. This appears to signal the proper scaling
behavior for these pion masses but only if we disregard the results for $M_{\pi}\lesssim 1\,\text{MeV}$. This might be helpful
information for a scaling analysis in lattice QCD simulations. However, we have to keep in mind that our results have been
obtained from a low-energy model in which the non-universal normalization constants 
are different from those found in QCD lattice simulations~\cite{Ejiri:2009ac,Karsch:2010ya}.

Finally we would like to comment on the pion mass dependence of the critical temperature in our model which is representative
for NJL/QM models. From the definition of the scaling variable $z$ and the universal peak position $z_{\rm p}\approx 1.3155$ 
of the rescaled susceptibility, we can estimate the dependence of the pseudo phase-transition temperature
$T_{\mathrm{p}}$ on the pion mass for small values of $H$, i. e. for small current quark masses. From Eq.~\eqref{eq:Mz} we deduce
\be
T_{\mathrm{p}}(H)\equiv T_{\mathrm{p}}(M_{\pi})=T_{\rm c} + T_0 z_{\rm p} \left(\frac{H}{H_0}\right)^{\frac{1}{\beta\delta}} = T_{\rm c} +  c_{\pi} M_{\pi}^{\frac{2}{\beta\delta}}\,,
\ee
where $c_{\pi}$ is a numerical constant which depends on the normalization constants and $T_{\rm p}(H\to 0)=T_{\rm c}$. 
Since $\beta\delta\approx 2$, we expect an almost linear dependence of $T_{\rm p}$ on $M_{\pi}$. This is in accordance with the findings from
lattice QCD simulations, see e.~g. Refs.~\cite{Karsch:2000kv,Karsch:2000zv}. However, the values for the non-universal 
normalization constants $T_0$ and $H_0$ differ significantly from those found in lattice simulations. As a consequence, the 
slope of $T_{\mathrm{p}}(M_{\pi})$ is about one order of magnitude larger in chiral low-energy models than in lattice simulations, 
see e.~g. \cite{Braun:2005fj} for a more quantitative
analysis of the slope. Thus, the universal properties of our model in the vicinity of the phase transition may agree with the
findings from lattice simulations, provided two-flavor QCD falls into the $O(4)$ universal class. 
Non-universal aspects such as the pion mass dependence of $T_{\rm p}$ seem to be incompatible with results from
full QCD. The discrepancy in the slope $c_{\pi}$ can be traced back to the parameter-fixing procedure in our model approach: We have fixed the Yukawa
coupling $g$, $m^2$ and $\lambda$ at the UV cutoff scale to reproduce the physical values for $f_{\pi}$, $M_{\pi}$ and $M_q$ for
a given value of $H$. For our studies with various pion masses we have then only varied $H$ but have left the UV values of
$g$, $m^2$ and $\lambda$ unchanged. However, these parameters have their own dependence on $H$ (i. e. on the current quark masses) which
is determined by quark-gluon dynamics at high momentum scales. We would like to add that the unknown dependence of the model parameters 
on the current quark masses may also affect the predictions of (P)NJL- and (P)QM-type models for the phase boundary at finite temperature and quark chemical 
potential~\cite{Fukushima:2003fw,Ratti:2005jh,Ghosh:2006qh,Sasaki:2006ww,Schaefer:2007pw,Cristoforetti:2010sn,Skokov:2010wb,Herbst:2010rf}.
This dependence cannot be computed with (P)NJL- and (P)QM-type models but is in fact accessible in RG studies of 
QCD~\cite{Gies:2002hq,Braun:2005uj,Braun:2006jd,Braun:2008pi, Braun:2009gm}. A determination
of the (current) quark mass dependence of these parameters from QCD RG flows is beyond the scope of 
this work and deferred to future work. 
\section{Scaling analysis in finite volume}
\label{sec:FV}
For systems in a finite volume, the critical behavior is modified because of the presence of the system boundaries. 
The linear extent of the volume $L$ appears as an additional relevant coupling: An actual critical point only exists 
in the limit $1/L \to 0$. Phase transitions and the associated singularities in the free energy appear only in this 
limit. However, even away from the critical point, the behavior of the system will still be controlled by the critical fixed point. 

The critical behavior is affected as soon as the correlation length is of the order of the extent of the volume, 
and consequently the scaling regime is characterized by the ratio of correlation length and volume size.

The singular part of the free energy density satisfies in leading order a scaling relation 
\be
f_s(t, h, L) =\ell^{-d}  f_s(t \ell^{y_t},h \ell^{y_h}, L \ell^{-1}),  
\ee
which now contains the volume extent $L$ as an additional coupling. By choosing the rescaling factor $\ell$ 
such that one argument is kept constant, we can use this relation to describe the system in terms of two variables 
only. In the limit of large volumes, the scaling behavior converges against the infinite-volume result. For this 
reason it is advantageous to choose the infinite-volume scaling variable $z$ as one of these two variables. In 
leading order, we can then express the behavior of the order parameter $M \equiv f_\pi$ as
\be
M(t, h, L) &=& L^{-\beta/\nu} Q_M(z, hL^{\beta \delta/\nu}), 
\ee
i.e. as a function of the scaling variables $z$ and $hL^{\beta \delta/\nu}$, where the volume dependence is 
now parameterized in the second variable. In order to reproduce the infinite-volume scaling relation in the limit 
$L \to \infty$, the finite-size scaling function must satisfy the relation
\be
\lim_{x \to \infty} Q_M(z, x) = x^{1/\delta} f_M(z)\qquad\text{with}\qquad x=hL^{\beta\delta/\nu}\,,
\label{eq:fssasymp}
\ee
where $f_M(z)$ is the infinite-volume scaling function from Eq.~\eqref{eq:Mz}.

A similar relation can be derived for the longitudinal susceptibility, where in leading order
\be
\chi_\sigma(t, h, l) &=& L^{\gamma/\nu} Q_\chi(z, hL^{\beta \delta/\nu}), 
\ee
and $Q_\chi(z, hL^{\beta\delta/\nu})$ is the finite-size scaling function for the susceptibility.
Because of the scaling relations between the critical exponents, $\gamma/\nu=(2 -\eta)$ is exactly $2$ 
for our results. For very small values of the symmetry-breaking parameter $h$, the scaling function 
$Q_\chi(z, hL^{\beta \delta/\nu})$ becomes essentially constant as a function of $hL^{\beta\delta/\nu}$ 
(see \cite{Braun:2008sg}). Consequently the susceptibility behaves as $\chi_\sigma \sim L^2$ for very 
small volumes. 

The finite-volume scaling behavior of $O(N)$ models has been considered in \cite{Engels:2001bq,Braun:2008sg}. 
As in the case of the infinite-volume scaling analysis, in these investigations relevant couplings which controlled the 
transition took the role of a temperature. In contrast, in the present investigation temperature is implemented as the 
finite extent of the Euclidean time axis in a four-dimensional Euclidean volume.

In $O(N)$ models as well as the chiral quark model considered here, at finite temperature the longitudinal correlation 
length is always bounded by the transverse correlation length. For this reason the scaling region 
can be characterized by the dimensionless product of box size and pion mass, $M_\pi L$. 

\begin{figure}[t]\centering
\includegraphics[scale=0.68]{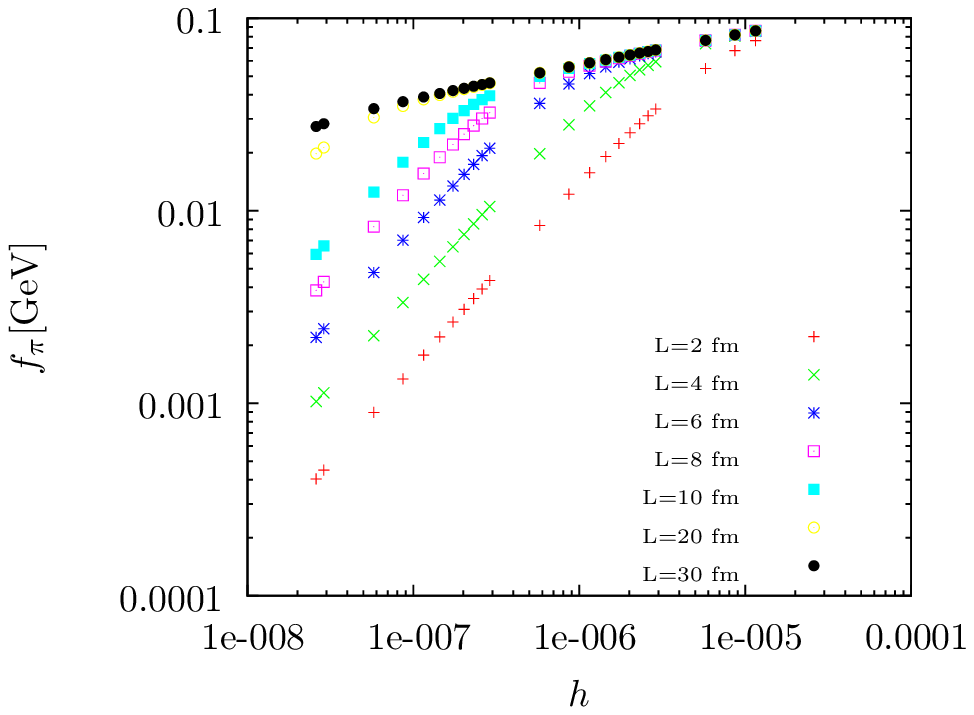}
\includegraphics[scale=0.68]{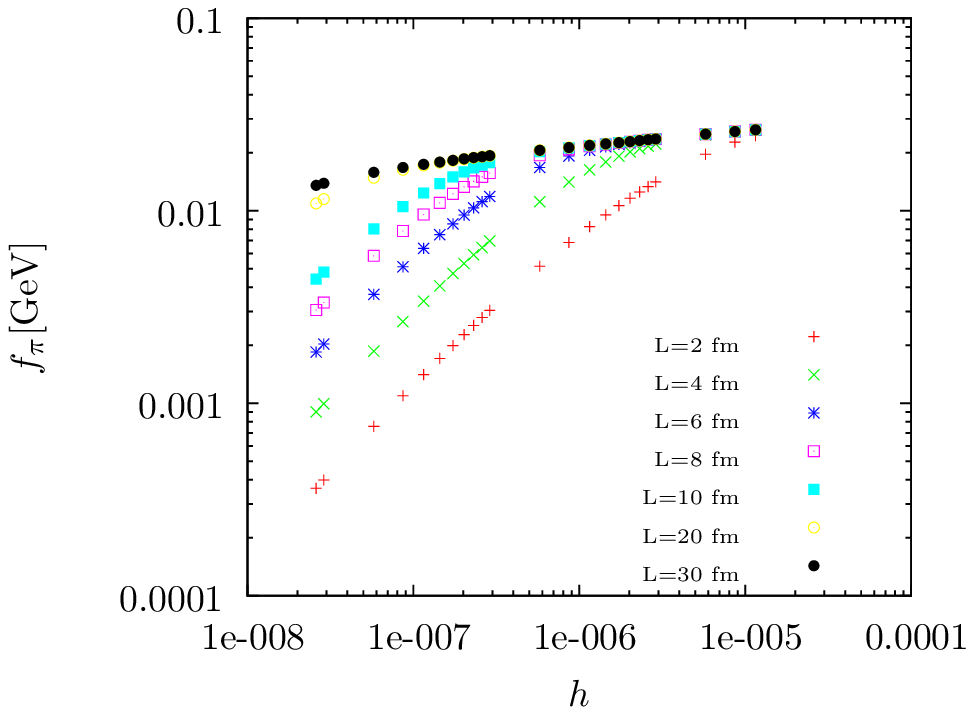}
\caption{Pion decay constant $f_{\pi}$  (order parameter) for $z=0$ (left panel) and $z=z_{\rm{p}}$ (right panel) as a 
function of the normalized symmetry breaking parameter $h$ for various values of the box length~$L$.
}
\label{fig:fsfpiz0}
\end{figure}

In Fig.~\ref{fig:fsfpiz0} we present our results for the order parameter $f_\pi$ in the quark-meson model 
in the finite-size scaling region. We have chosen to present results at fixed values of the scaling variable $z$. 
The choice $z=0$ corresponds to the critical temperature $T=T_c$, and $z=z_{\mathrm{p}}=1.3155$ corresponds
to the position of the maximum (peak) in the longitudinal susceptibility $\chi_\sigma$, which are shown in the left and right panel of Fig.~\ref{fig:fsfpiz0}.

In agreement with our expectations \cite{Engels:2001bq,Braun:2008sg}, we find strong finite-volume effects for 
small values of the symmetry breaking, and a convergence to the infinite-volume behavior for strong symmetry 
breaking in the raw (unscaled) results shown in Fig.~\ref{fig:fsfpiz0}. We calculated the order parameter as a 
function of the symmetry-breaking parameter $h$ for different values of the volume size, from $L=2$ fm to $L=30$ fm. 
For large values of the symmetry breaking the correlation length is small compared to the volume size, and the 
results converge to the infinite-volume behavior for large $h$.  For decreasing $h$, the correlation length 
increases, and depending on the volume size $L$ the deviation from the infinite-volume behavior sets in when the order 
of the correlation length approaches the volume size. For smaller volume size $L$, this happens for larger values of $h$.

\begin{figure}[t]\centering
\includegraphics[scale=0.68]{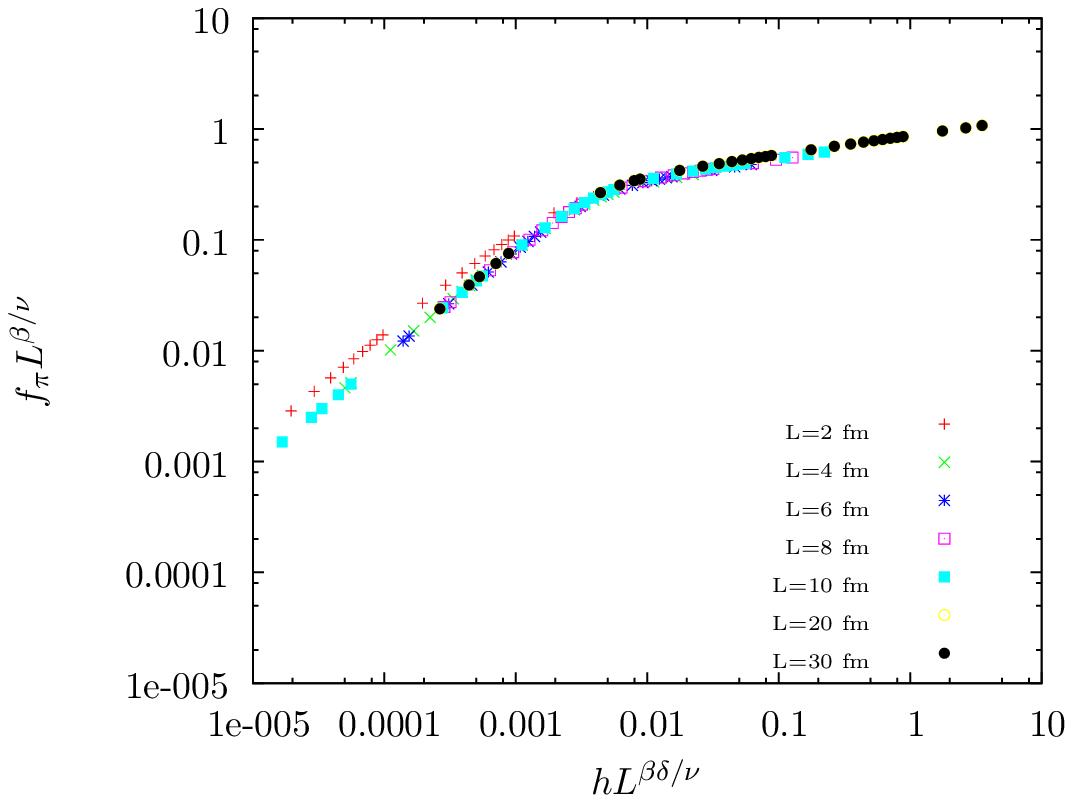}
\includegraphics[scale=0.68]{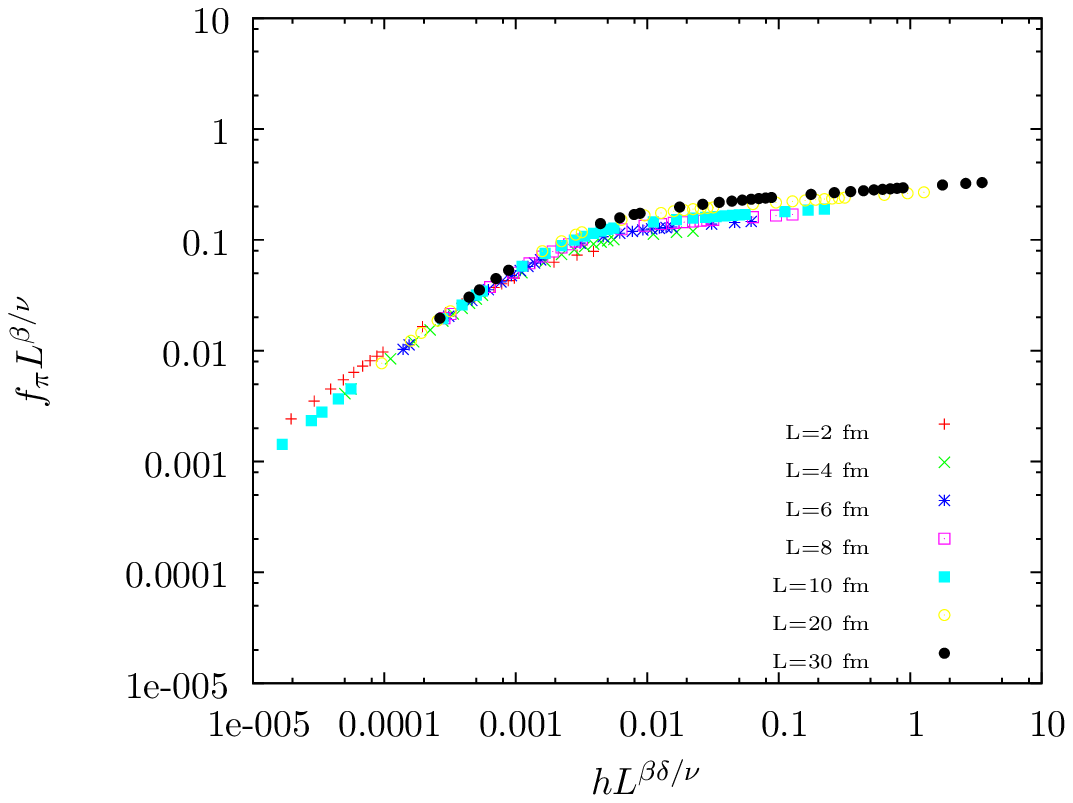}
\caption{Finite-size scaled order parameter $f_{\pi} L^{\beta/\nu}$ for $z=0$ (left panel) and $z=z_{p}$ (right panel) as a 
function of the scaling variable $hL^{\beta\delta/\nu}$ for various values of the box length $L$.}
\label{fig:fsscaledfpiz0}
\end{figure}

In Fig.~\ref{fig:fsscaledfpiz0} the results for the finite-size scaled order parameter 
$L^{\beta/\nu} M \equiv L^{\beta /\nu} f_\pi$ are shown as a function of the finite-size scaling variable 
$hL^{\beta\delta/\nu}$, for both $z=0$ and $z=z_{\mathrm{p}}$ (left and right panel).  The rescaled results fall 
onto a single scaling curve and thus show the expected finite-size scaling behavior. We can distinguish two 
different regimes in the rescaled results, where the rescaled results follow different power laws. 

For small volume size compared to the correlation length, the behavior is dominated by the effects of the finite 
volume. This part of the scaling function corresponds to those parts of the curves in Fig.~\ref{fig:fsfpiz0} that 
deviate from the infinite-volume behavior. The "bend", where the slope in the double-logarithmic plot changes, 
characterizes the region in $\xi/L$ where deviations from the infinite-volume behavior become the dominant effects.

To give some guidance about the finite-size scaling region, in Tab.~\ref{tab:bend} we list the pion masses which 
correspond to the value of the symmetry-breaking parameter at the position where the slope changes. For a range 
of volume sizes, we give the value $M_\pi(T\to 0, L\to \infty)$ of the pion mass at zero temperature and in infinite 
volume for the value of $h$ at the bend point. We list in this example the values at the critical temperature, 
i.e. for the choice $z=0$.
\begin{table}
\begin{tabular}{lrrrrrrr}
\hline\hline
$L$ [fm] & 
\phantom{0}2 & 
\phantom{0}4 & 
\phantom{0}6 &
\phantom{0}8 &
10 &
20 &
30 \\\hline
$M_\pi$ [MeV] &
\phantom{X}308&
\phantom{X}139&
\phantom{XX}85&
\phantom{XX}60&
\phantom{XX}45&
\phantom{XX}19&
\phantom{XX}11\\
\hline
$M_\pi L$ & 
3.12&
2.82&
2.59&
2.43&
2.30&
1.94&
1.75\\
\hline
\hline
\end{tabular}
\caption{Pion mass $M_\pi(T\to 0, L\to \infty)$ corresponding to the value of the scaling variable $hL^{\beta\delta/\nu}$ 
at the bend point of the scaling curve in Fig.~\ref{fig:fsscaledfpiz0}, and corresponding values of the dimensionless product $M_\pi L$. 
The values given are for the results with $z=0$.}
\label{tab:bend}
\end{table}

For large values of the finite-size scaling variable $hL^{\beta\delta/\nu}$, i.e. for small $\xi/L$, the curve is characterized by the infinite-volume behavior. 
Due to the asymptotic large-volume behavior Eq.~\eqref{eq:fssasymp}, the finite-size scaling function for the order parameter behaves as $x^{1/\delta}$ for 
large values of the scaling variable $x=hL^{\beta \delta/\nu}$. In the double-logarithmic plot in Fig.~\ref{fig:fsscaledfpiz0}, 
the exponent appears as the slope for large values of the scaling variable. We have checked explicitly that the inverse slope 
agrees with the value for the critical exponent $\delta$ from our analysis, and find very good agreement for the results with $z=0$.

In the results from the $O(4)$ model presented in \cite{Braun:2008sg}, the asymptotic behavior for large volumes follows a power 
law with an exponent close to $1/\delta$ for both $z=0$ and $z=z_{\mathrm{p}}$ as expected, with a slightly larger exponent for 
$z=z_{\mathrm{p}}$ than for $z=0$. In contrast, in the present case the exponent and thus the slope of the curve in the 
double-logarithmic plot for $z=z_{\mathrm{p}}$ is somewhat smaller than for $z=0$. The agreement between the finite-size 
scaled results for different values of $L$ is better for $z=0$ that for $z=z_{\mathrm{p}}$, which was also observed for the 
lattice spin model in \cite{Engels:2001bq}. The difference in the behavior of the asymptotic scaling function between the 
quark-meson model in the present paper and the $O(4)$ model in \cite{Braun:2008sg} is likely due to quark effects, as we will 
discuss below.

As discussed in \cite{Braun:2008sg}, the presence of the additional coupling $L$ requires the determination of an additional 
non-universal normalization constant $L_0$ in order to determine a truly universal scaling function. A possible normalization 
condition is to require $\xi = L_0 t^{-\nu}$ for $h=0$. A direct comparison between the different scaling functions would 
be possible only after such an additional normalization.

\begin{figure}[t]\centering
\includegraphics[scale=0.68]{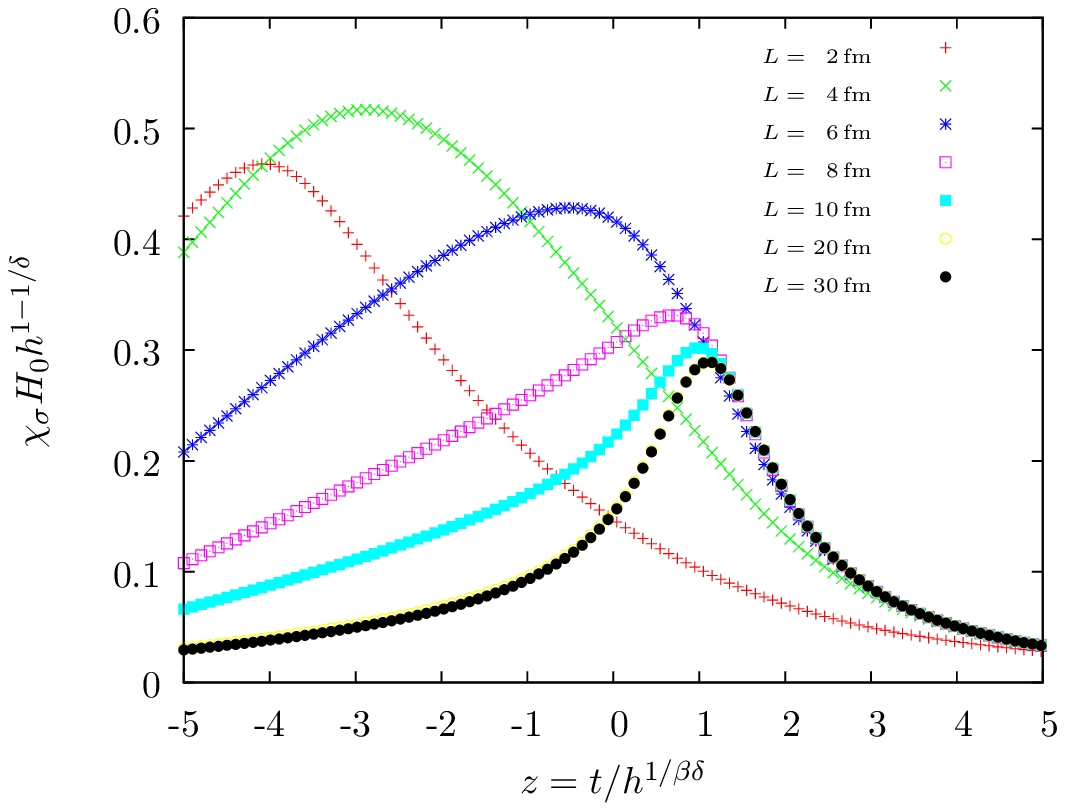}
\includegraphics[scale=0.68]{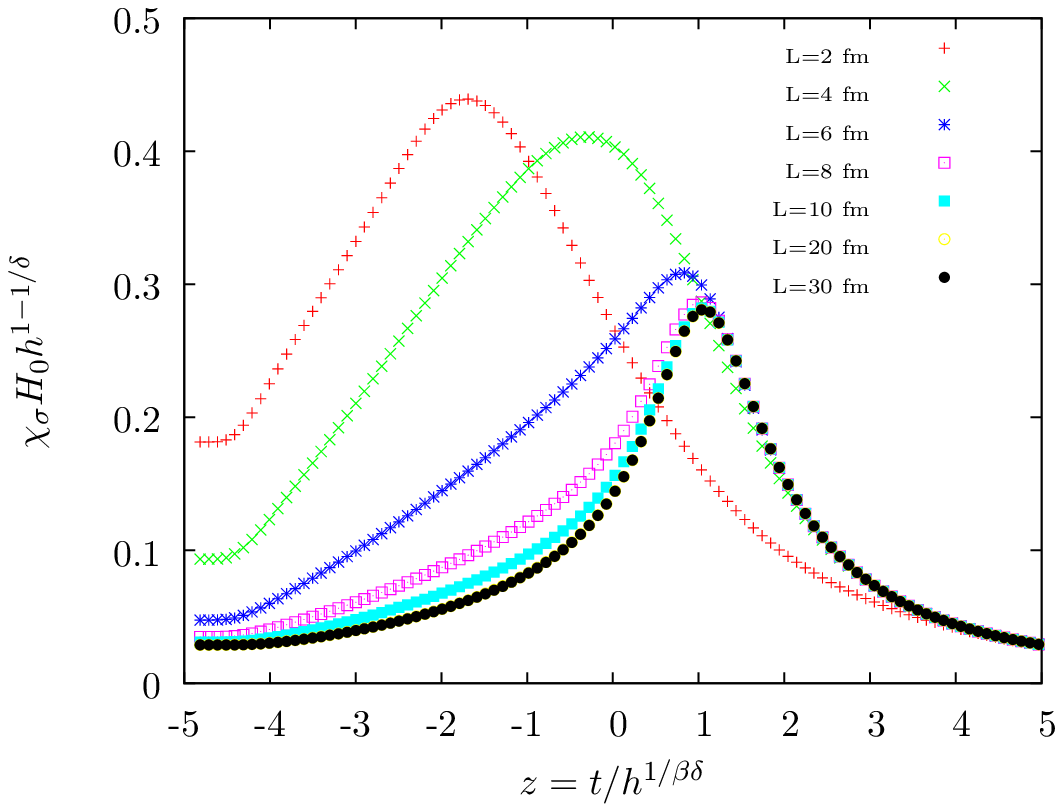}
\caption{Rescaled $\sigma$-susceptibility for $M_{\pi}= 48\,\text{MeV}$ (left panel) and $M_{\pi}= 75\,\text{MeV}$ 
(right panel) as a function of $z$ for various box lengths $L$.}
\label{fig:susc_finiteL_mpi48}
\end{figure}

\begin{figure}[t]\centering
\includegraphics[scale=0.68]{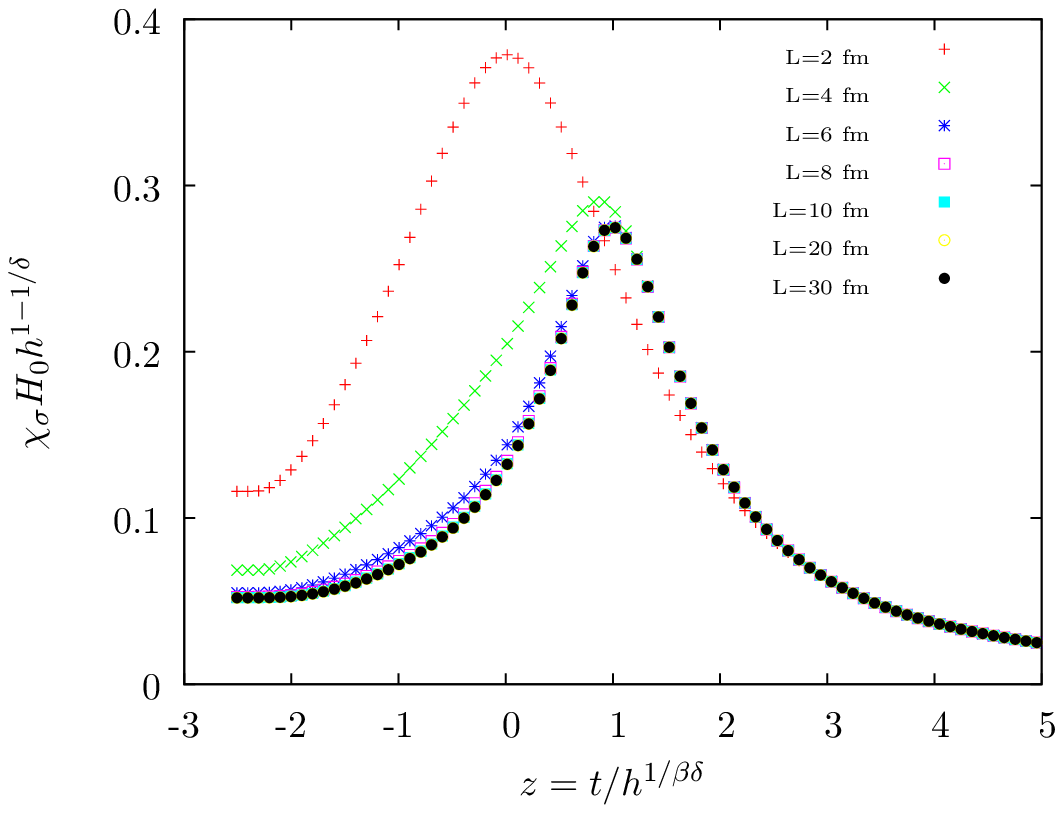}
\includegraphics[scale=0.68]{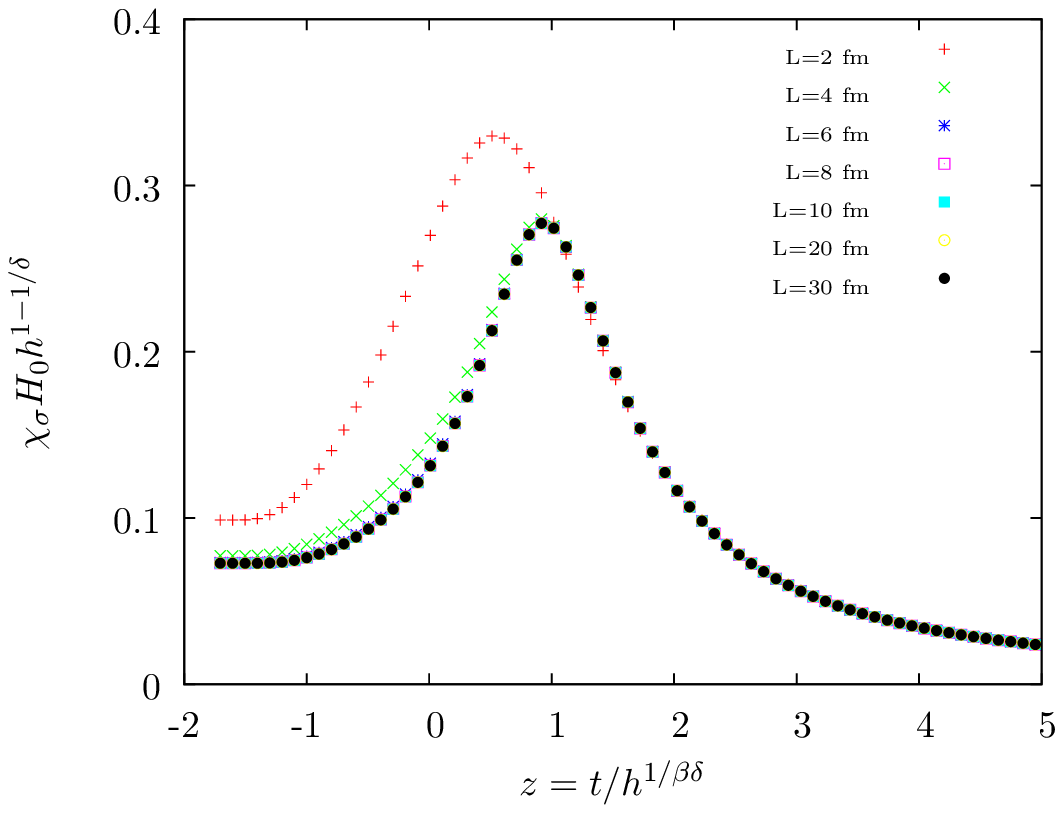}
\caption{Rescaled $\sigma$-susceptibility for $M_{\pi}= 138\,\text{MeV}$ (left panel) and $M_{\pi}= 200\,\text{MeV}$ 
(right panel) as a function of $z$ for various box lengths $L$.
}
\label{fig:susc_finiteL_mpi138}
\end{figure}

We will now turn from the dedicated finite-size scaling analysis to the finite-volume effects that appear if one performs a 
conventional infinite-volume scaling analysis in this volume and pion mass region.
In Figs.~\ref{fig:susc_finiteL_mpi48} and~\ref{fig:susc_finiteL_mpi138} we  show results for the susceptibility in a 
finite volume for fixed values of the symmetry breaking parameter $h$, i.e. for fixed values of the pion mass $M_\pi$. 
As examples, we have chosen the values $M_\pi = 48$ MeV, $M_\pi = 75$ MeV, $M_\pi=138$ MeV, and $M_\pi = 200$ MeV. The parameters for 
the model are chosen in such a way that these values for the pion mass are obtained in the limit $L \to \infty$ and $T \to 0$. 

We present the results for the rescaled susceptibility $\chi_\sigma H_0 h^{1-1/\delta}$ as a function of the 
infinite-volume scaling variable $z=t/h^{1/\beta\delta}$, as one would do in the absence of finite-volume effects. 
The purpose of this exercise is to investigate the deviations from the infinite-volume scaling behavior due to such 
effects and to estimate the volume size where they become relevant. This volume size is of course strongly dependent 
on the pion mass.

As discussed above, for small volume size the susceptibility scales as $\chi_\sigma \sim L^2$ since the correlation 
length is bounded in a finite volume, $\xi \lesssim L$. Whether this behavior can be observed in the presence of explicit 
symmetry breaking depends once again on the dimensionless product $M_\pi L$. For large symmetry breaking ($M_\pi L \gg 1$), 
this scaling behavior will not be evident. 

Starting with a comparatively small pion mass of $M_\pi = 48$ MeV (left panel of Fig.~\ref{fig:susc_finiteL_mpi48}), we find that the 
expected decrease in the susceptibility with the volume can be observed only between the smallest volumes with $L=2$ fm and $L=4$ fm. 
This conforms to a rough estimate from the bound that the mass of the Goldstone mode places on the correlation length: a pion mass 
$M_\pi = 48$ MeV corresponds to a length scale of approximately $4$ fm. For larger volumes, the susceptibility is still bounded by the 
pion mass, i.e. the amount of explicit symmetry breaking. The volume size becomes the limit only below this size.

Nevertheless, we observe significant finite-volume effects in the susceptibility for larger volumes and pion masses. From Figs.~\ref{fig:susc_finiteL_mpi48} 
and \ref{fig:susc_finiteL_mpi138}, we see that for a comparatively large pion mass of $M_\pi=200$ MeV significant deviations from the infinite-volume 
scaling appear only for the smallest volume with $L=2$ fm. For a physical pion mass of $M_\pi=138$ MeV, these deviations appear below $4$ fm, 
and for $M_\pi = 75$ MeV for volume sizes below $L= 6$ to $8$ fm.     

This behavior, however, is very different from the naive expectation for the finite-volume behavior: The susceptibility increases 
in these volumes and is larger than in the infinite-volume limit, which cannot be explained in terms of a simple cutoff effect for 
the long-range fluctuations.

We interpret this behavior as a quark effect in the chiral quark-meson model. This behavior is specific to the choice of periodic boundary conditions for the 
quark fields in the spatial directions of the finite volume. As observed in a systematic study of the effects of the quark boundary conditions in the 
quark-meson model \cite{Braun:2005gy}, zero-mode effects in a finite volume lead to an increase in the chiral quark condensate and a 
corresponding decrease of the pion mass in an intermediate volume size. This affects in turn also the longitudinal susceptibility $\chi_\sigma$ 
and leads to larger values than in the infinite-volume limit. In contrast, with a choice of anti-periodic boundary conditions for the quark fields, 
the lowest momentum mode acts as a mass gap, which increases with decreasing volume, and these effects are absent~\cite{Braun:2005gy}. 

It is likely that these quark effects are stronger in the model calculation than in actual QCD, where quarks are subject to confinement at low momentum scales.
Nonetheless, evidence for this effect in QCD comes from lattice QCD simulations and the approach via Dyson-Schwinger Equations: 
A decrease of the pion mass in a finite volume was observed in quenched \cite{Guagnelli:2004ww} as well as unquenched \cite{Orth:2005kq} 
lattice simulations, and in a Dyson-Schwinger approach to QCD~\cite{Luecker:2009bs}. The authors of the last reference interpret their results as an effect of quenching, 
but cannot exclude that it is also present if all pion effects are taken into account.

Our results demonstrate that a finite-size scaling analysis is feasible for sufficiently small volumes. On the other hand, 
additional quark effects already manifest themselves already for larger volume sizes and affect the analysis in terms of 
infinite-volume scaling analysis, if they are not taken into account.

\section{Conclusions}\label{sec:conc}
In this paper we have performed a scaling analysis of the chiral order parameter in infinite and finite volume with the aid of an 
effective low-energy model, namely the chiral quark-meson model. Using functional RG methods we have computed the
scaling functions in infinite and finite volume and have analyzed the behavior of the chiral susceptibility in detail. In contrast
to earlier studies of the scaling behavior of three-dimensional theories, we have performed the scaling analysis in $d=4$ Euclidean dimensions,
where the temperature in a field-theoretical sense is determined by the extent of the Euclidean volume in (imaginary) time
direction. With a such a setup, scaling of the order parameter associated with a three-dimensional universal class is only
observed when the conditions for dimensional reduction are met. We have confirmed that the observation of scaling 
requires therefore that $T_{\rm c} L\gg 1$, $M_{\pi}L\gg 1$ and $M_{\pi}/T_{\rm c}\ll 1$, where $T_{\rm c}$ is the critical 
temperature for $M_{\pi}\to 0$ and $L\to\infty$.

In our study of scaling in infinite volume we have only observed scaling behavior of the chiral susceptibility and the chiral order 
parameter for rather small pion masses, $M_{\pi} \lesssim 1\,\text{MeV}$. Corrections to scaling become soon apparent for 
$M_{\pi} > 1\,\text{MeV}$. Moreover, we have compared our result for the scaling function of the chiral order parameter for 
$M_{\pi}\to 0$ with the rescaled order parameter for different pion masses with $M_{\pi}\gtrsim 75\,\text{MeV}$. 
We have found that the results for the
rescaled order parameter for these pion masses fall almost on one line. This appears to indicate the proper scaling behavior in
this pion mass regime but only if we disregard the results for $M_{\pi}\ll 1\,\text{MeV}$. This observation might be of interest for scaling
studies in lattice QCD simulations, even though non-universal quantities such as the normalization constants $T_0$
and $H_0$ in our scaling analysis differ from those in full QCD. Turning the argument around, this observation and, in
consequence, the too-strong dependence of the critical temperature on the pion mass might also be considered as a warning for
studies of the phase-diagram based on (P)NJL/(P)QM-type models. As we have argued, this discrepancy can possibly be traced back
to the unknown dependence of the (UV) parameters on the current quark mass in these models. One way to lift this 
ambiguity in the parameters is the use of QCD RG flows~\cite{Gies:2002hq,Braun:2005uj,Braun:2006jd,Braun:2008pi, Braun:2009gm,Braun:2009ns}.

In finite volumes we have also found that the chiral order parameter shows the expected scaling behavior. 
In addition, we have studied the finite-volume
effects which appear if a conventional infinite-volume scaling analysis is performed in finite volumes with box lengths $L=2,\dots,30\,\text{fm}$
and with pion masses $M_{\pi}\gtrsim 48\,\text{MeV}$. This allows us to depict clearly the deviations from infinite-volume scaling
behavior which are due to the presence of a finite volume. We have found that the chiral susceptibility scales as $\chi_{\sigma}\sim L^2$ for
small volumes, $M_{\pi}L < 1$. If this condition is not met, then the susceptibility does not scale as $\chi_{\sigma}\sim L^2$. On the 
contrary, we have shown that the height of the peak of the susceptibility increases in this pion mass regime with decreasing volume size.
We would like to stress that this is a quark effect which might be less pronounced in lattice QCD simulations, but which is
presumably still present. It has indeed been found in lattice simulations~\cite{Guagnelli:2004ww}, DSE studies~\cite{Luecker:2009bs} and in an
earlier RG study~\cite{Braun:2005fj} that the pion mass exhibits a minimum as a function of the volume size depending on the 
actual value of the dimensionless quantity $TL$. This effect can be traced back to the spatial zero modes of the fermions
and it is therefore only present when periodic boundary conditions for the fermions are applied~\cite{Braun:2005fj}.
In any case, this volume dependence of the pion mass clearly affects the susceptibility.

Overall, we believe that our results provide further insight into the various aspects of the scaling behavior of the chiral order parameter and 
will provide helpful information for the scaling analysis of lattice QCD results. In addition, we think our results will also help to further 
improve the construction of QCD low-energy models.

\begin{acknowledgments}
The authors gratefully acknowledge useful discussions with Holger Gies, Frithjof Karsch, Jan M. Pawlowski, Wolfram Weise 
and Andreas Wipf. This work was supported by the Research Cluster "Structure and Origin of the Universe" and the DFG Research Training 
Group "Quantum and Gravitational Fields" ({\mbox GRK 1523/1}).
\end{acknowledgments}

\bibliography{QMscaling}

\end{document}